\documentclass[pre,aps,twocolumn,showpacs,superscriptaddress,floatfix]{revtex4}

\usepackage{amssymb}
\usepackage{graphicx}
\usepackage{amsmath}
\usepackage{latexsym}
\usepackage{verbatim}

\begin{document}

\title{LinkRank: Finding communities in directed networks}

\author{Youngdo Kim}
\author{Seung-Woo Son}%
\affiliation{
Department of Physics, Korea Advanced Institute of Science and Technology, Daejeon 305-701, Korea.}

\author{Hawoong Jeong}
 \email{hjeong@kaist.edu}
\affiliation{
Department of Physics, Korea Advanced Institute of Science and Technology, Daejeon 305-701, Korea.}
\affiliation{
Institute for the BioCentury, Korea Advanced Institute of Science and Technology, Daejeon 305-701, Korea.}
\date{\today}

\begin{abstract}
To identify communities in directed networks, we propose a generalized form of
modularity in directed networks by introducing a new quantity LinkRank, which can be considered as the PageRank of links. This generalization is consistent with the original
modularity in undirected networks and the modularity optimization methods
developed for undirected networks can be directly
applied to directed networks by optimizing our
new modularity. Also, a model network, which can be used as a benchmark network in further community studies, is proposed to verify our method. Our method is supposed to find communities effectively in citation- or reference-based directed networks.
\end{abstract}


\maketitle

\section{introduction}
Uncovering the structure of nature is an essential part of our effort to understand the world around us~\cite{Kemp2008}. It is the same when it comes to the complex network~\cite{Albert2002,Newman2003a,Dorogovtsev2001a}, which is considered as a simple but powerful representation of real-world complex systems. Among the underlying structures of complex networks, community structure is considered to be important since it has proven to be strongly related to the dynamics and functions of complex networks~\cite{Colizza2006a,Guimera2006}. Hence, considerable attention has been given from various fields to uncover the community structure of networks~\cite{Fortunato2007a,Porter2009}.

Generally, a community is a group of nodes in which the nodes are densely inter-connected compared to the rest of the network. And, a network is considered to have community structure when there are more links placed within the communities and fewer links placed between the communities. Uncovering the community structure in a given network means finding the best community assignment describing the underlying community structure well. In order to decide which community assignment is better than any other possible assignments, a benefit function is required. Modularity, which was proposed by Newman and Girvan~\cite{Newman2004}, is one of the most widely used benefit functions. Although it has been reported that there exist the resolution limit~\cite{Fortunato2006b} and the bias towards balanced partitions~\cite{Delvenne2008,Lambiotte2008c}, modularity is still considered to be an efficient measure of uncovering the community structure.

Even after the modularity is chosen as the benefit function, there still lies a difficult problem. Finding the community assignment with the highest modularity is not an easy task as the exhaustive optimization of modularity is usually impossible. In order to overcome this difficulty, many methods~\cite{Danon2005,Newman2006} have been proposed to obtain the best approximation of the highest modularity in a reasonable time, and most of those methods are working effectively compared to the computing power they require. It is important to notice that those methods can only be applied to undirected networks, of which links have no specific direction, because the definition of modularity is limited to undirected ones. However, many complex networks in the real-world are directed ones, such as the World Wide Web, citation networks, phone call or email networks, etc.

In many directed networks, the direction of a link contains important information such as asymmetric influence or information flow. A link between a pair of nodes may represent a fundamentally different dynamics when its direction is reversed. Any kind of approach that disregards the direction of links may fail to understand the dynamics and the function of directed networks. Also, any kind of community finding approaches may fail to detect the communities correctly if the direction of the link is not considered properly. Then, there lies the fundamental question of the problem of community identification in directed networks: How should the direction of links be considered? This is a question that is not only essential to the community identification but also important to the fundamental understanding of the directed networks.

Several recent studies~\cite{Newman2006b,Guimera2007,Rosvall2007,Leicht2007a} have tried to answer this question. However, it is important to notice that the listed methods do not share a common definition of the community structure in directed networks. The method of Newman and Leicht~\cite{Newman2006b}, and the method of Guimera \textit{et al.}~\cite{Guimera2007} have the similar definition by which nodes are assigned to the same community when the nodes are linked to similar neighbors. The definition of community used in those works is different from the general definition of community. A fundamentally different approach is adopted in the work of Rosvall and Bergstrom~\cite{Rosvall2007}. They used an information theory based method that also does not seem related to the modularity optimization method. Leicht and Newman~\cite{Leicht2007a} proposed a method that is different from those previous ones. They adapted a generalized modularity~\cite{Arenas2007} to identify the community structure in directed networks. Since the generalized modularity is consistent with the original modularity in undirected networks, the advantage of this method is apparent: the modularity optimizing algorithms developed in undirected networks are supposed to be applicable to directed networks by this method. However, we find that there may exist some limitations in this method. In Sec.~\ref{sec:genmod}, it is shown that the generalized modularity may not work as they described in Ref.~\cite{Leicht2007a}, and an alternative meaning of the generalized modularity will be discussed.

In this paper we propose a new generalization of modularity based on LinkRank, which is a quantity indicating the importance of links in directed networks. The definition of community is also changed according to this new modularity. It will be shown that this definition consists well with the old definition of community and it considers the links of different direction properly. The application to a model network in Sec.~\ref{sec:application} shows that our method works effectively in detecting communities. We deal with weighted networks in the derivations, since binary networks can be considered a special kind of weighted networks in which the weight of all links is one.

\section{Generalized Modularity
\label{sec:genmod}}

In undirected networks, a well established method to find communities is the modularity optimizing method, which is finding a good community assignment
of networks which maximizes the benefit function named modularity
$Q^{ud}$~\cite{Newman2004,Newman2004a}. The modularity is defined as
\begin{equation}
Q^{ud} = {\frac{1}{2M}} \sum_{i,j}{\biggl[ w_{ij}-\frac{w_iw_j}{2M}\biggr]\delta_{c_i,c_j}},
\label{eq:ModUnd}
\end{equation}
where $w_{ij}$ is an element of the weighted adjacency matrix which represents the weight of the link
between node $i$ and $j$, $w_{i}=\sum_{j}{w_{ij}}$ is the strength of node
$i$, and the total strength is $2M=\sum_{i}{w_i} = \sum_{i}\sum_{j}{w_{ij}}$.

The modularity can be understood as the difference of two quantities. The
first one is the fraction of links within communities, and the second one is the
expected value of the first one in a network with the same community
divisions and the same strength sequence but randomly connected links. Modularity
$Q^{ud}$ approaches $1$ when a strong community structure is found and approaches $0$
when the fraction of links within the communities is no better than a random case.
However, this does not mean the maximized modularity of every random network is around zero. Some random networks may have very high maximized modularity due to fluctuations in the establishment of links~\cite{Guimera2004}.

Arenas \textit{et al.}~\cite{Arenas2007} proposed a generalization of
modularity in directed networks by simply replacing the strength terms into directional ones. The generalized modularity can be described as
\begin{equation}
Q^{d} = {\frac{1}{M}} \sum_{i,j}{\biggl[ w_{ij}-\frac{w_i^{out}w_j^{in}}{M}\biggr]\delta_{c_i,c_j}},
\end{equation}
where $w_{ij}$ represents the weight of link pointing from node $i$ to node $j$,
$w_i^{out}=\sum_{j}{w_{ij}}$ and $w_j^{in}=\sum_{i}{w_{ij}}$ are respectively
the out-strength and in-strength of node $i$ and node $j$, and the total strength is
$M=\sum_{i}{w_i^{out}} =\sum_{j}{w_j^{in}}= \sum_{i,j}{w_{ij}}$.

\begin{figure}
\includegraphics[width=0.9\columnwidth]{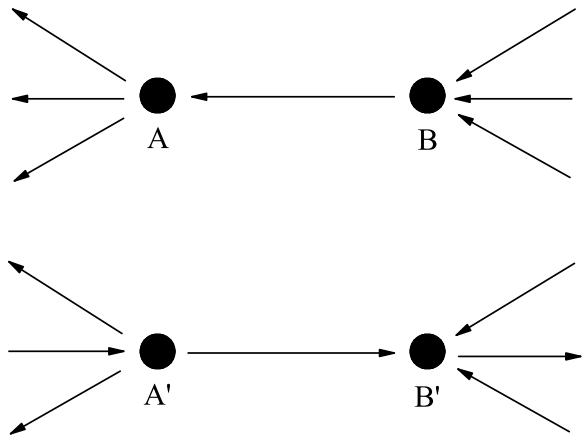}
\caption{The generalized modularity does not distinguish the direction of
links. Node $A$, $B$, $A'$, $B'$ are four nodes in a directed binary network.
The out-strength and in-strength of those nodes are
$w_A^{out}=w_{A'}^{out}=w_B^{in}=w_{B'}^{in}=3$, and
$w_A^{in}=w_{A'}^{in}=w_B^{out}=w_{B'}^{out}=1$. The contribution of the link
between node $A$ and node $B$ is equal to the contribution of link between node
$A'$ and node $B'$: $q_{AB}=q_{A'B'}=1-{5}/{M}$.
}
\label{fig:4nodes}
\end{figure}

Leicht and Newman~\cite{Leicht2007a} used this new definition of modularity to
find communities in directed networks, both for computer-simulated networks and
real-world networks. They described the meaning of this generalized modularity
as follows. For a pair of nodes labeled $A$ and $B$, when node $A$ has high out-degree
and low in-degree while $B$ has the reverse, then a directed link connecting $A$
and $B$ is more likely to point from $A$ to $B$ than the opposite direction.
Hence, if a directed link running from $B$ to $A$ is found in a network, it is a bigger surprise than a link from $A$ to $B$. The link from $B$ to $A$ should contribute more to the modularity since modularity should be high for statistically surprising configurations.

However, the generalized modularity may not work as
described above. Because $\delta_{c_i,c_j}$ is equal to $\delta_{c_j,c_i}$, the
generalized modularity $Q^d$ is able to be derived as
\begin{equation}
Q^{d} = {\frac{1}{2M}} \sum_{i,j}{\biggl[
w_{ij}+w_{ji}-\frac{w_i^{out}w_j^{in}}{M}-\frac{w_j^{out}w_i^{in}}{M}\biggr]\delta_{c_i,c_j}},
\end{equation}
Considering $w_{ij}$ and $w_{ji}$ are summed together and controlled by the same $\delta_{c_i,c_j}$,
it is questionable that the generalized modularity can distinguish  the direction of links.
Fig.~\ref{fig:4nodes} represents a part of a directed binary network. Node $A$ and $A'$ have
a higher out-degree, while node $B$ and $B'$ have a higher in-degree. According
to Leicht and Newman's explanation, node $A$ and $B$ should be more likely
to be divided into the same community than node $A'$ and $B'$.
However, the contribution of both pairs to the generalized modularity are
actually equal:
\begin{eqnarray}
q_{AB}=0+1-\frac{3\times 3}{M}-\frac{1\times 1}{M}=1-\frac{5}{M},\\
q_{A'B'}= 1+0-\frac{3\times 3}{M}-\frac{1\times 1}{M}=1-\frac{5}{M},
\end{eqnarray}
where $q_{ij}\equiv
w_{ij}+w_{ji}-{w_i^{out}w_j^{in}}/{M}-{w_j^{out}w_i^{in}}/{M}$ is the contribution of the link between node $i$ and $j$ to
the generalized modularity $Q^{d}$. Therefore, it is doubtful that the generalized modularity work as described above. Then, there may arise the following questions: How could
the generalized modularity identify communities in
directed networks? And what is the meaning of the generalized modularity?

As explained in the appendix of Ref.~\cite{Arenas2007}, the relation between the generalized
modularity $Q^{d}$ in directed networks and the modularity $Q^{ud}$ in
undirected networks can be expressed as
\begin{equation}
Q^{d}=
Q^{ud}+\frac{1}{4M^2}\sum_{ij}{\Delta_{i}\Delta_{j}}\delta_{{c_i},
{c_j}},
\label{eq:derv}
\end{equation}
where $Q^{ud}$ is the modularity of the undirected network
which is generated from the original
directed network by ignoring link directions, and $\Delta_i=w_i^{out}-w_i^{in}$ is the \textit{net-strength} of node $i$.

Hence, the second term in Eq.~(\ref{eq:derv}) should be the additional
information considered in the generalized modularity $Q^{d}$. It is easy to
notice that the second term would give a positive contribution to $Q^{d}$
only when $\Delta_i$ and $\Delta_i$ are both positive or
both negative. Therefore, the effect of the second term
is putting together a group of nodes that have positive net-strength, and another group of nodes having negative net-strength. In an extreme case, where there is no community structure when the link directions are ignored, the first term $Q^{ud}$ in Eq.~(\ref{eq:derv}) makes no contribution to the modularity and the second term would contribute importantly to the modularity.
It is obvious that the directed network in this case should be divided into two communities by maximizing the generalized modularity $Q^{d}$; the one community contains nodes with positive net-strength and the other community contains nodes with negative net-strength. Example networks presented in Ref.~\cite{Leicht2007a} are similar to the extreme
case described above, in which no communities can be found if the directions of links are ignored. Those networks were divided into two communities: a community composed of nodes with positive net-strength and another community composed of nodes with negative net-strength. No effect, such as nodes connected by a link of surprising direction are more likely to be in the same community, has been considered in this approach.

\section{LinkRank and a new generalization of modularity
\label{sec:linkrank}}

The most important property of the directed network is definitely the direction of links. For example, in a directed network of webpages, a webpage
with more incoming hyperlinks is much more important and more likely being
visited than a webpage with more outgoing hyperlinks, even if those two pages
have the same degree, which is the sum of in-degree and out-degree.
Furthermore, a webpage linked by another important webpage should be more
important than a webpage linked by a minor webpage. Therefore, a link from an
important page should be more important than a link from a minor page, i.e. a
link from an important page should be more likely to be an intra-community link. If
one wants to identify the communities in a directed network, it is
necessary to take into account this unique property of the directed networks.
Actually, there already exists a quantity called PageRank that exploits this unique property in directed networks.

PageRank~\cite{Brin1998,Langville2006} is an analysis algorithm used by Google to rank
the webpages in the World Wide Web, which is a typical directed network. PageRank
assigns a quantity that indicates the
importance of a webpage with the thesis that a webpage is
important if it is pointed to by other important pages. Mathematically, PageRank
is the probability of a particular page being visited by a random surfer who
clicks the hyperlinks in webpages randomly. The PageRank equation can be
described as
\begin{equation}
\boldsymbol{\pi}^{T}=\boldsymbol{\pi}^{T}\boldsymbol{G},
\label{eq:PageRank}
\end{equation}
where $\boldsymbol{\pi}^T$ is the stationary row vector of $\boldsymbol{G}$ called
the PageRank vector, and each element $\pi_i$ is the probability that a random
walker is going to visit the node $i$ in the
stationary state. $\boldsymbol{G}$ is called the Google Matrix and it is the
probability matrix for the random walk process. Each element $G_{ij}$ is the
probability that a random walker on node $i$ moves toward node $j$ in the next random walk step.
$G_{ij}$ is defined as $G_{ij}=w_{ij}/w_i^{out}$, where $w_{ij}$ is the element of the
weighted adjacency matrix in directed networks and $w_i^{out}$ is the out-strength of node
$i$.

In a directed network, there may exist some dangling nodes, which is a node with only incoming links, and ``trap region'', which is a region where the random walker can only move in but not move out. In this case, the Google Matrix defined as above cannot guarantee the existence of
the stationary row vector $\boldsymbol{\pi}^T$, because $\boldsymbol{G}$ may not satisfy the requirements of the stochastic matrix in
the Markov process~\cite{Meyer2000}. To avoid this problem, the Google Matrix is
actually defined as
\begin{equation}
G_{ij}=\alpha \frac{w_{ij}}{w_i^{out}} + \frac{1}{N}(\alpha a_i +1-\alpha),
\label{eq:GoogleMatrix}
\end{equation}
where ($1-\alpha$) is the teleportation probability, by which the random walker
stops following the hyperlinks and opens a random webpage, and $a_i$ is equal to
one only if node $i$ is a dangling node; otherwise $a_i$ is zero. The value of $w_{ij}/w_i^{out}$ is set to $0$ when $w_i^{out}=0$.
By adding $a_i$ and $\alpha$ to the definition of $\boldsymbol{G}$, the random walker would not be trapped in any part of the network during the random walk process. Mathematically speaking, the purpose of this modification is
to make the Google Matrix $\boldsymbol{G}$ a
completely dense, stochastic, and
primitive matrix. Therefore, there always exists a stationary vector
$\boldsymbol{\pi}^T$
for the Google Matrix $\boldsymbol{G}$~\cite{Langville2006,Meyer2000}.

Following the idea of PageRank, we propose a concept of LinkRank, which
indicates the importance of the links instead of the importance of nodes in
PageRank. Similar to the definition of PageRank, LinkRank
of a particular link should be equal to the probability that a random walker
follows the link from node $i$ to node $j$ in the stationary state. With the
definition of $\pi_i$ and $G_{ij}$, LinkRank can be simply defined as
\begin{equation}
L_{ij}=\pi_i G_{ij},
\label{eq:LinkRank}
\end{equation}
where $\pi_i$ is the $i$th element of PageRank vector $\boldsymbol{\pi}$, and $G_{ij}$ is
the element of Google Matrix $\boldsymbol{G}$.

As described in Sec.~\ref{sec:genmod}, the modularity in undirected networks is qualitatively defined as
\begin{eqnarray}
Q^{ud} &=&\textrm{(fraction of links within communities)} \nonumber\\
&& - \textrm{(expected value of this
fraction)},
\label{eq:OldModularity}
\end{eqnarray}
where the expected value is calculated in a network with the same community
divisions and the same strength sequence but randomly connected links. In this
paper, we propose a new definition of modularity for both directed and
undirected networks as
\begin{eqnarray}
Q^{lr} &=&\textrm{(fraction of time spent walking within}\nonumber\\
&& \textrm{communities by a
random walker)} \nonumber \\
&&-\textrm{(expected value of this fraction)}.
\label{eq:NewModularity1}
\end{eqnarray}
Reminding that modularity defines intrinsically communities, it is important to notice that the definition of community is changed in our method. According to the new modularity $Q^{lr}$, a community is no longer a group of nodes in which links are more densely located. Instead, a community is a group of nodes in which a random walker is more likely to stay. Although this definition seems out of nowhere, it will be shown in the following part that this definition is consistent with the old one in the undirected networks and considers the links of different direction properly.

By using LinkRank, this new definition can be written in a mathematical form as
\begin{equation}
Q^{lr} = \sum_{i,j}{L_{ij}}\delta_{c_ic_j} - \sum_{i,j}{\textrm{E}(L_{ij})}\delta_{c_ic_j},
\label{eq:NewModularity2}
\end{equation}
where $\textrm{E}(L_{ij})$ is the expected value of $L_{ij}$ in the null model. In
Eq.~(\ref{eq:NewModularity2}), it is easy to notice that the first term is the fraction of time spent on
walking within communities by a random walker since $L_{ij}$ is the probability of the random walker
following the link from $i$ to $j$, and the second term is the
expected value of this fraction. Both terms correspond to the first and
second terms in Eq.~(\ref{eq:NewModularity1}) respectively.

In order to calculate the expected value of $L_{ij}$, a null model has to be chosen first. In the definition of modularity in undirected networks, the standard null model is chosen as a network that has the same strength (or degree in binary networks) sequence as the original network (i.e. the expected strength of each node is conserved and the links are randomly rewired). In directed networks, however, it is not proper to choose the same null model as in undirected networks since strength is not directly related to the random walk process. Instead, PageRank is the intrinsic property of nodes through the random walk process. It can be shown that the null model of conserving strength sequence does not detect communities as we expected, while the null model of conserving PageRank sequence does~\cite{YKim2009a}. Therefore, we choose a random network, in which the PageRank sequence is conserved and the links are randomly rewired, as the null model to compare with.

In this null model, the expected value of $L_{ij}$ can be calculated as follows. As defined above, LinkRank $L_{ij}$ is the probability that a random walker is moving from node $i$ to $j$ in the stationary state, and PageRank $\pi_i$ is the probability that a random walker is visiting node $i$ in the stationary state. In order to move from node $i$ to $j$, the random walker would have to visit node $i$ in the previous step and to visit node $j$ in the next step. The probability of visiting node $i$ is $\pi_i$, and the probability of visiting node $j$ in the next step is $\pi_j$ because the connection between node $i$ and $j$ in the original network is not conserved in the null model. Therefore, the probability that a random walker moving from node $i$ to $j$ in the null model is $\pi_{i}\pi_j$, which means the expected value of $L_{ij}$ in the null model is
\begin{equation}
\textrm{E}(L_{ij})=\pi_{i}\pi_j.
\end{equation}
Finally, the modularity in directed networks is
\begin{equation}
Q^{lr} =
\sum_{i,j}\biggl[{L_{ij}- \pi_i \pi_j} \biggr]\delta_{c_ic_j}.
\label{eq:NewModExp1}
\end{equation}

Interestingly, our new definition of modularity consists well with the old definition of modularity in
Eq.~(\ref{eq:OldModularity}). It is well known that when the random teleportation is not considered, the PageRank vector
$\boldsymbol{\pi}^{ud}$ in undirected networks satisfies
$\pi_i^{ud}=w_i/2M$,
where $w_i$ is the strength of node $i$, and $2M=\sum_{i}{w_i}$ is the total strength
of the undirected network~\cite{Meyer2000,JDNoh2004b,Perra2008,Masuda2008a}.
This means that the probability that a random walker visiting node $i$ in the
stationary state is only related to the local structure of node $i$, instead of
being related to the global structure.
Because every link in the undirected network is a bidirectional path, there are
no dangling nodes or trap regions in undirected networks. Therefore,
the second term of Eq.~(\ref{eq:GoogleMatrix}) can also be ignored and
the Google Matrix $\boldsymbol{G}^{ud}$ of the undirected network is $G_{ij}=w_{ij}/w_i$.
Then the LinkRank of the undirected network is
\begin{equation}
L_{ij}^{ud}=\pi_i^{ud}G_{ij}^{ud}=\frac{w_{ij}}{2M}.
\label{eq:LinkRankUndirected}
\end{equation}
The expected value of LinkRank in the undirected network is
\begin{equation}
\textrm{E}^{ud}(L_{ij})=\pi_i^{ud} \pi_j^{ud} = \frac{w_i}{2M}\frac{w_j}{2M}.
\end{equation}
Then, the undirected version of our new modularity is
\begin{equation}
Q^{lr} =
\sum_{i,j}{\biggl[\frac{w_{ij}}{2M}-\frac{w_i}{2M}\frac{w_j}{2M}\biggr]\delta_{c_i
c_j}},
\label{eq:ExpectedLinkRankUndirected}
\end{equation}
which is identical to the definition of modularity in Eq.~(\ref{eq:ModUnd}). This means that our new definition of community---a community is a group of nodes in which a random walker is more likely to be trapped in---is consistent with the old definition of community---a community is a group of nodes in which links are more densely located. 

Also, the new modularity in Eq.~(\ref{eq:NewModExp1}) has a similar form with the one-step stability in Delvenne \textit{et al.}~\cite{Delvenne2008}, which is another work revealing the connection between random walk and the modularity.

An remarkable advantage of our method
is that all the established optimizing techniques~\cite{Danon2005,Newman2006} developed to maximize the old
modularity in undirected networks can be applied to our method directly, except
a few algorithms in which some adjustments are needed.
For example, to apply the eigenvector-based method~\cite{Newman2006}
, a small trick introduced
in Ref.~\cite{Leicht2007a} is needed to restore the symmetry of
the modularity matrix.

It is important to notice that our method cannot be applied to all kinds of networks because the direction of a link does not have a universal meaning in all kinds of directed networks. For example, although a word adjacency network~\cite{Newman2006b} and the World Wide Web are both directed networks, the meaning of direction is fundamentally different in those two networks. The direction of link in a word adjacency network describes the relative position of the linked words in a sentence, while the direction of link in the World Wide Web indicates the citation or reference. Information can spread by following the directed links in the latter case. As PageRank can be applied to any collection of entities linked with citation and reference, our method is supposed to be able to detect communities in directed networks based on citation and reference. This does not mean that our method is limited to linked documents only. As social networks such as directed friendship networks, phone call networks and email networks can be considered as a general form of citation/reference networks, our method could be used to detect communities in those networks too.

The parameter $\alpha$ controls the priority given to the network structure as opposed to the teleportation effect. When $\alpha$ is close to $1$, the random walk process would be more dependent on the network structure. Therefore, the PageRank, LinkRank and the new modularity would be more likely to capture the characteristics of the network if $\alpha$ is closer to $1$. However, it has been reported that PageRank becomes more sensitive to the slight change of network structure when $\alpha$ gets closer to $1$~\cite{Langville2006}. Considering that it is impossible for a network to describe the underlying system fully and correctly, PageRank cannot capture the characteristics of the network when it is too sensitive to the slight change of the structure. A balance has to be taken between respecting the network structure and reducing sensitivity. The choice of $\alpha$ is an important issue to any other random work research on directed networks. Thus, the research on the effect of $\alpha$ requires a general discussion over random walk problems in directed networks. We would like to tackle this problem in future studies, but it is beyond the scope of this paper. In current status, $\alpha=0.85$ seems a good choice since it is widely used by other researchers~\cite{Brin1998,Langville2006,Perra2008}.

\section{Relation with other community identification works in
directed networks}

\begin{figure}
\includegraphics[width=0.9\columnwidth]{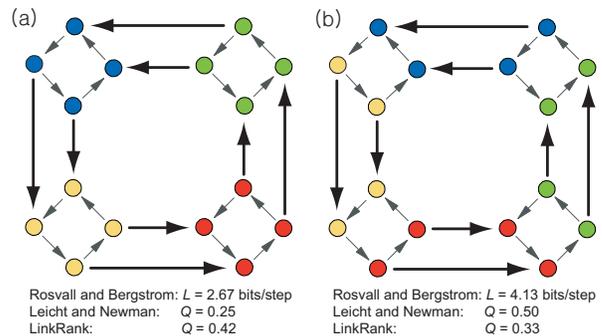}
\caption{(color online) Maximizing our new modularity finds the same community assignment as the
method proposed by Rosvall and Bergstrom does. The weight of the bold links is
twice the weight of normal links, and the color of a node indicates the
community that the node belongs to. (a) Community assignment given
by optimizing the new modularity of our method or by optimizing the map equation
of Rosvall and Bergstrom~\cite{Rosvall2007}. The new modularity for this community assignment
is $Q^{lr}=0.42$, while the modularity used by Leicht and Newman~\cite{Leicht2007a} is $Q^d=0.25$.
(b) Community assignment given by optimizing the modularity that
is used by Leicht and Newman. Our new modularity for this assignment is $Q^{lr}=0.33$, while the modularity used by Leicht and Newman
is $Q^d=0.50$.
}
\label{fig:16nodes}
\end{figure}

As described in Sec.~\ref{sec:genmod}, Leicht and Newman~\cite{Leicht2007a} proposed that the links with opposite directions should be considered differently to identify communities correctly in directed networks. The way the direction information is considered in our new modularity is very
similar to the way considered in Leicht and Newman's work. Let's consider the pair of nodes $A$ and $B$ again. When node $A$ has lower PageRank and node $B$ has higher PageRank, the LinkRank of the link pointing from $A$ to $B$
is more likely to be lower than the LinkRank of the link from $B$ to $A$.
Therefore, node $A$ and node $B$ are more likely to be in the same community if
the link is pointing from node $B$ to node $A$ than if the link is pointing the opposite direction. Thus,
the asymmetry effect of the link direction, which Leicht and Newman wanted to
include in their method, is well considered in our new definition of
modularity in a quite systematic way by applying the theory of random walk.

Also, the work of
Rosvall and Bergstrom~\cite{Rosvall2007} is directly related to our work. In their work, they proposed an
information theory-based method to detect communities in directed networks. This
method can be briefly described as follows. For a particular community assignment
of a directed network, a node name is assigned to each node in the network and
a community name is assigned to each community in the network. The nodes in the
same community should have different names to distinguish with, and the nodes
from different communities may share the same names because they can be
distinguished by their community names. Given the names of nodes and
communities, a description can be assigned to a trajectory of a random walk on
the network. The description records the name of each node being visited, and the
name of the community which the currently visiting node belonging to is recorded
before the node name only when the random walker is coming from
a node which belongs to another community. Thus, this description
is unique to each trajectory of random walk. When a random walker is more likely to stay within a group of nodes than average, dividing this group of nodes into the same community will make the description shorter. Therefore, the community structure can be identified by minimizing the length of trajectory description.

Although our method seems to have no relevance with this method, both methods share the same definition of the community
structure---a community is a group of nodes that a random walker is
more likely to be trapped in instead of moving out of the group in a few steps.
The simple directed network composed of sixteen nodes in Fig.~{\ref{fig:16nodes}} was originally proposed in the work
of Rosvall and Bergstrom. The weight of the bold links is twice the weight
of the other links. As shown in the figure, the communities detected by our method
are identical to the results of Rosvall and Bergstrom, preferring the
configuration with long persistence time. The new modularity for the
community assignment in
Fig.~{\ref{fig:16nodes}} (a) is much larger than the modularity in
Fig.~{\ref{fig:16nodes}} (b), while the modularity calculated by the
method of Leicht and Newman has a higher value for the community assignment in Fig.~{\ref{fig:16nodes}} (b),

\section{Application to a model network}
\label{sec:application}

\begin{figure}
\includegraphics[width=0.9\columnwidth]{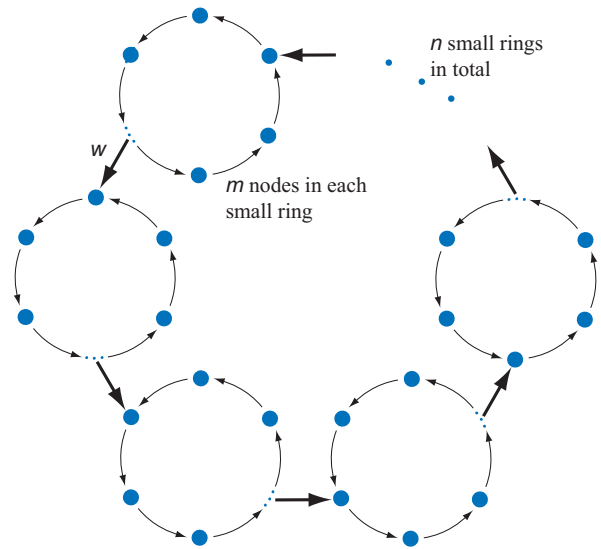}
\caption{A model network that is designed to verify our method. This network is a directed network composed of $n$ sub- networks, and the sub-networks are embedded on a ring structure. Also, each sub-network is a small ring composed of $m$ nodes. Each small ring has an entrance node, which is the node receiving a directed link from upper stream ring, and an exit node, which is giving a directed link to the down stream ring, and the entrance node and the exit node are placed at the opposite side of each other. The direction is chosen counterclockwise both in the small ring and in the big ring. The weight of link between sub-networks is a tunable parameter $w$, while the weight of link in every sub-network is fixed as $1$. According to our definition to community in directed networks, each small ring should be considered as a community as long as $w$ is not significantly large because the random walker would be more likely to be trapped in each small ring rather than freely moving between the small rings.
}
\label{fig:GroupOfRings}
\end{figure}

\begin{figure}
\includegraphics[width=0.9\columnwidth]{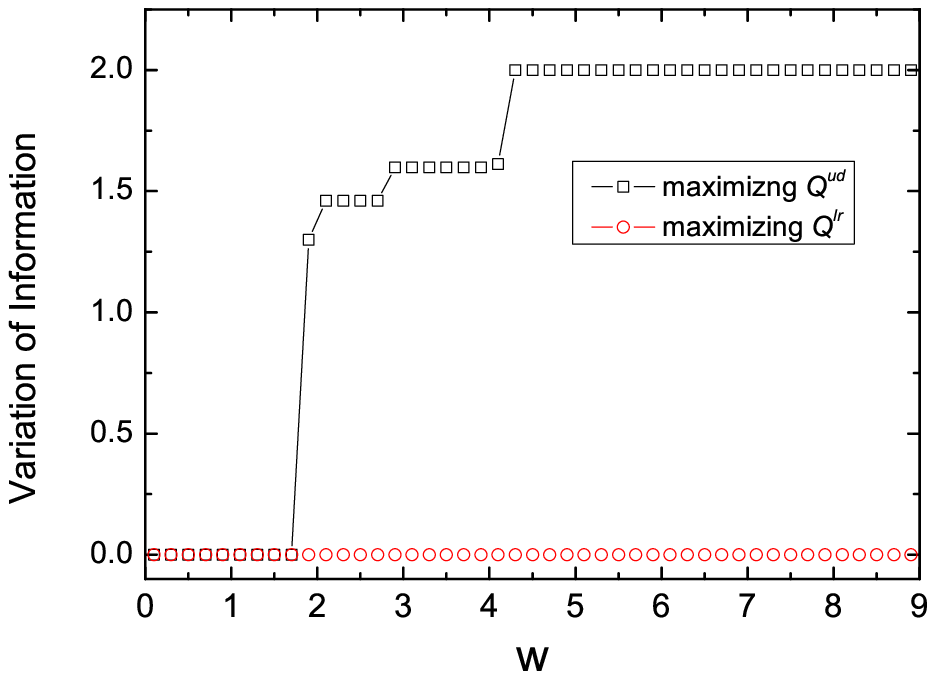}
\includegraphics[width=0.9\columnwidth]{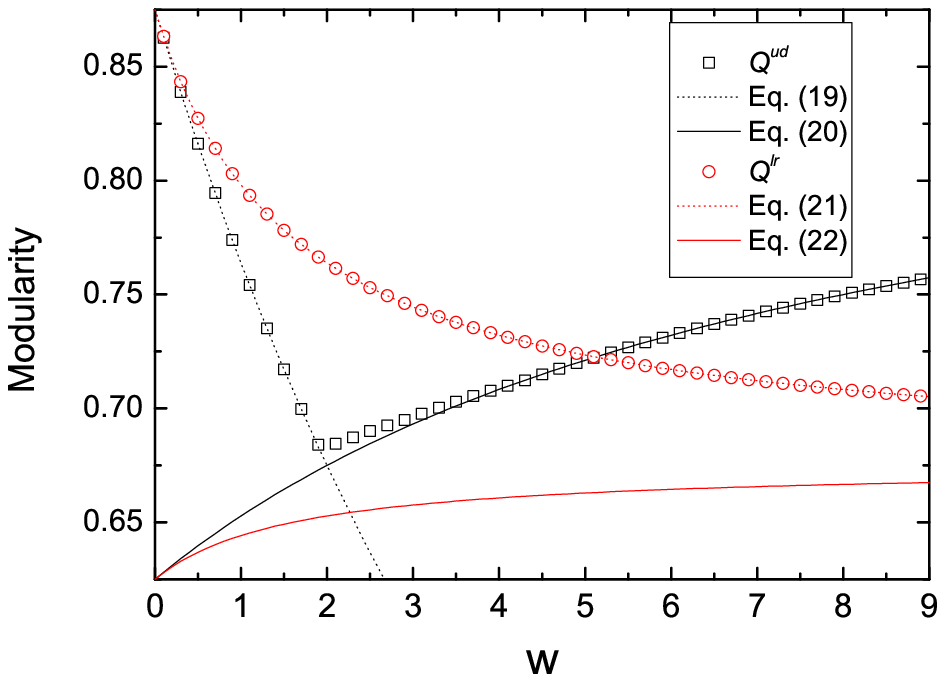}
\caption{(color online) Variation of information and modularity as the function of $w$ for a model network of $m=n=8$. (a) Square (circle) symbols represent the VOI between the embedded community structure and the community assignment given by optimizing $Q^{ud}$ ($Q^{lr}$). (b) Square symbols and circle symbols represent the highest $Q^{ud}$ and the highest $Q^{lr}$ found by the simulated annealing algorithm during the community identifying process of both methods. The black dotted line, the black solid line, the red dotted line and the red solid line are corresponding to the functions in Eq.~(\ref{eq:Qa}), Eq.~(\ref{eq:Qb}), Eq.~(\ref{eq:Qc}) and Eq.~(\ref{eq:Qc}) separately.
}
\label{fig:alphaVQ}
\end{figure}

The network illustrated in Fig.~{\ref{fig:GroupOfRings}} is a directed model network that is designed to verify our method. In this network, $n$ directed small rings are embedded on a big ring and each small ring is composed of $m$ nodes. The weight of the links between small rings is a tunable parameter $w$, while the weight of other links is fixed as $1$. The small rings are the embedded communities of this model network. When $w$ is small, it will be difficult for the random walker to escape from each small ring, and the communities should perfectly overlap with the small rings, remembering that we consider the community in directed networks as a group of nodes where a random walker is more likely to be trapped. And when $w$ gets larger, it will become easier for the random walker to move out of each small ring, and consequently, the embedded community structure would be more difficult to be identified. When $w$ is large enough, it will be not reasonable to identify the small rings as communities.

If the directions of links are ignored, a random walker would be more likely to move out of each small ring than in the case when link directions are considered. Therefore, a direction-ignoring method would not effectively detect community structure of the model network while our direction-considering method can detect community structure correctly when $w$ is neither too small nor too large. To prove this, we have to quantitatively compare the identified community assignments with the embedded community structure. Here, we use the variation of information (VOI), which is described as a true metric of the community assignment by Karrer \textit{et al.}~\cite{Karrer2007}, to compare the different community assignments. The VOI is defined as
\begin{equation}
V(A,B) = -\sum_{i=1}^{C^A}\sum_{j=1}^{C^B}\biggl({\frac{n_{ij}^{AB}}{N} \log_2{ \frac{n_{ij}^{AB}}{n_j^B}}+{\frac{n_{ij}^{AB}}{N} \log_2{ \frac{n_{ij}^{AB}}{n_i^A}}}\biggr)},
\label{eq:VAB}
\end{equation}
where $A$ and $B$ are the two community assignments to be compared with, $C^A$ and $C^B$ are the total number of communities of assignment $A$ and $B$ correspondingly, $N$ is total number of nodes, $n_i^A$ is the number of nodes in $i$th community of assignment $A$, $n_j^B$ is the number of nodes in $j$th community of assignment $B$, and $n_{ij}^{AB}$ is the number of nodes which are in $i$th community of assignment $A$ and in $j$th community of assignment $B$ at the same time. Generally, the VOI is large if the compared two community assignments are significantly different, and it is small when the community assignments are similar.

We tested both methods in a model network of $64$ nodes, in which $m=8$ and $n=8$. Since there are no dangling nodes or trap regions in our model network, the teleportation rate $(1-\alpha)$ is taken as zero. The community assignments detected by both methods are compared with the embedded community structure, and the difference is measured by the VOI. The results of VOI and modularity are plotted in Fig.~{\ref{fig:alphaVQ}}. Relatively small $m$ and $n$ are chosen in order to correctly find the highest modularity, and simulated annealing algorithm~\cite{Guimera2005c}, which is an algorithm showing best performance in the benchmark~\cite{Danon2005}, is chosen as the optimizing algorithm.

As illustrated in Fig.~{\ref{fig:alphaVQ}}(a), when $w$ is small, the community assignments detected by both methods are identical to the embedded structure. However, when $w$ is larger than $1.9$, the VOI for the direction-ignoring method starts to get a non-zero value and becomes larger until it fixes at a stationary value. This means that the community assignment detected by this method is getting more and more different from the embedded community structure and finally fixes at a stationary configuration. Meanwhile, the community assignment detected by our method is identical to the embedded community structure in the illustrated range of $w$. This significant difference indicates that a method that ignores the direction of links cannot identify the community structure effectively in this model network, while our method can detect the communities effectively.

\begin{figure}
\centering
\includegraphics[width=0.9\columnwidth]{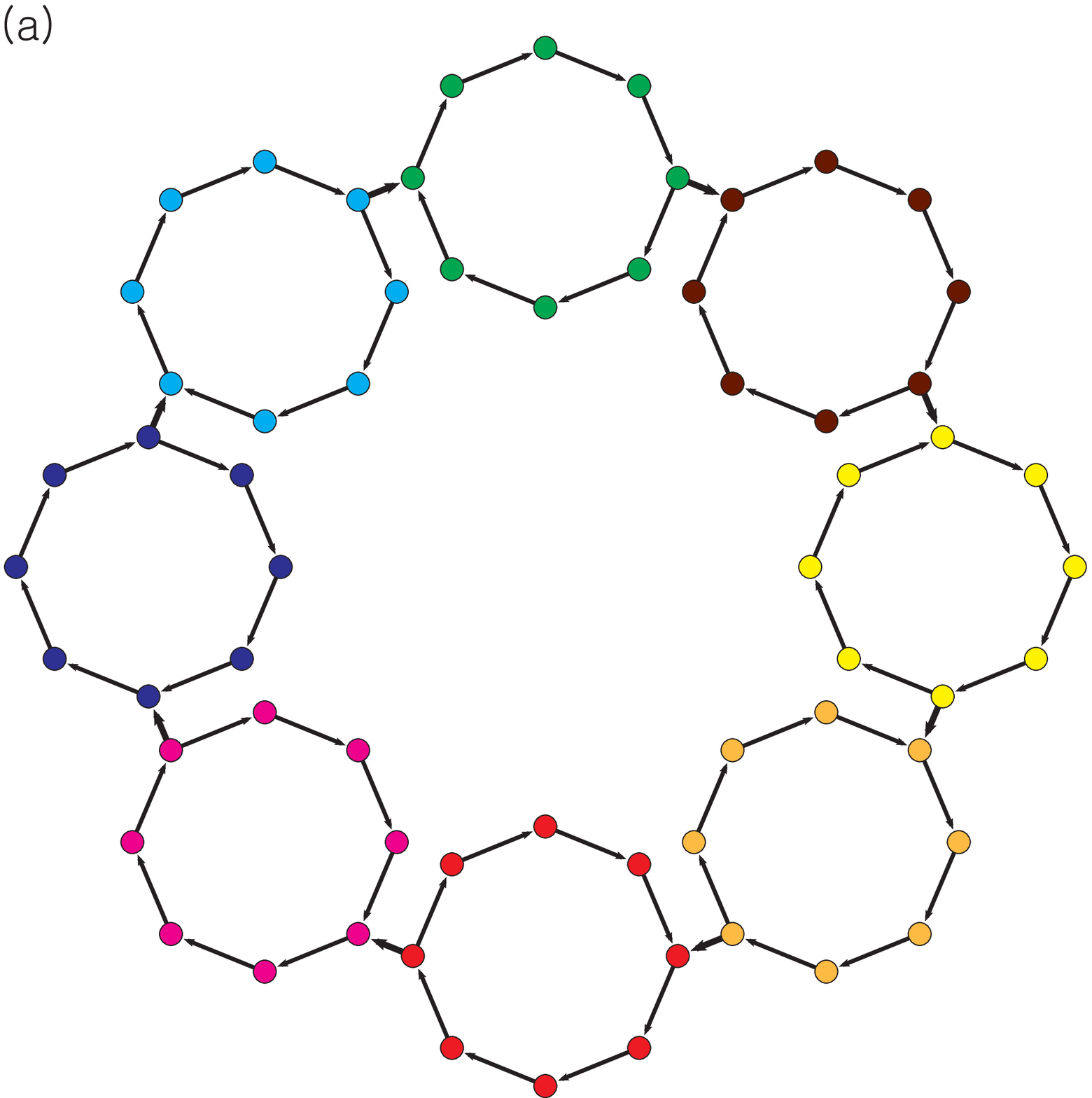}
\includegraphics[width=0.9\columnwidth]{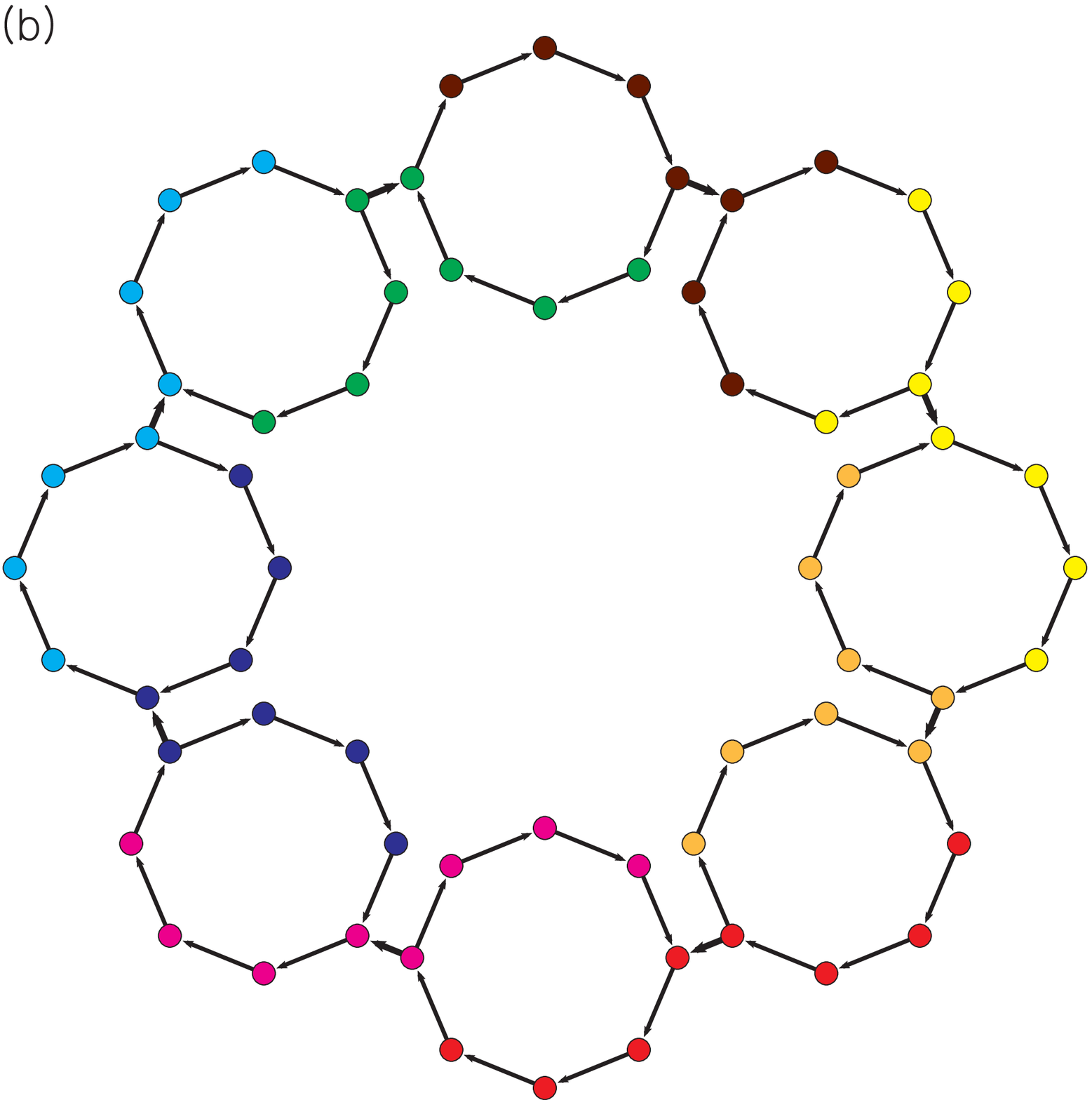}
\caption{(color online) Community assignments given by direction-ignoring method and our method in the model network of $m=n=8$. (a) The community assignment given by direction-ignoring method, when $w$ is smaller than $1.9$. And the community assignment given by our method through the illustrated values of $w$. (b) The stationary community assignment given by direction-ignoring method, when $w$ is larger than $4.3$.}
\label{fig:Configurations}
\end{figure}

Further investigation on the values of modularity and the corresponding community assignments aid the better understanding of this model network and our method. Fig.~{\ref{fig:alphaVQ}}(b) shows that the modularity values given by both methods also show different behaviors as the weight $w$ becomes larger. When $w$ is small, the community assignment detected by the direction-ignoring method is identical to the embedded community structure. The modularity for this community assignment can be expressed analytically as
\begin{equation}
Q^{ud}_{A} =\frac{m}{m+w}-\frac{1}{n},
\label{eq:Qa}
\end{equation}
and the black dotted line in Fig.~{\ref{fig:alphaVQ}}(b) shows the curve of this function. When $w$ is larger than $4.3$, a stationary community assignment emerges. This community assignment is illustrated in Fig.~{\ref{fig:Configurations}}(b). This result is easy to understand since the weight $w$ is large now and the nodes connected by the inter-ring links are more likely to be assigned into the same community. The modularity for this community assignment is
\begin{equation}
Q^{ud}_{B}= \frac{m+w-2}{m+w} - \frac{1}{n},
\label{eq:Qb}
\end{equation}
and the black solid line in Fig.~{\ref{fig:alphaVQ}}(b) shows the curve of this function. In the range of $w\in(1.9,4.3)$, the modularity values given by simulated annealing are slightly larger than the values given by Eq.~(\ref{eq:Qb}), because there exist some transitional community assignments.

Similar analysis can also be performed to our method. For the community assignment that is identical to the embedded community structure (Fig.~{\ref{fig:Configurations}}(a)), the new modularity given by our method is
\begin{equation}
Q^{lr}_{A}=\frac{mw+2m}{mw+2m+2w} - \frac{1}{n}.
\label{eq:Qc}
\end{equation}
For the community assignment of Fig.~{\ref{fig:Configurations}}(b), which is the stationary community assignment given by the direction-ignoring method for large $w$, the new modularity is
\begin{equation}
Q^{lr}_{B}=\frac{mw+2m-4}{mw+2m+2w} - \frac{1}{n}.
\label{eq:Qd}
\end{equation}
It is easy to notice that $Q^{lr}_{A}$ is always larger than $Q^{lr}_{B}$ no matter what value $w$ takes. This means that no matter how large $w$ is, a community assignment as Fig.~{\ref{fig:Configurations}}(b) will never be detected by our method.

In Fig.~{\ref{fig:alphaVQ}}(b), both the results of simulated annealing algorithm and the analytical functions indicate that, as the weight $w$ increases, $Q^{ud}$ is decreasing rapidly while $Q^{lr}$ is decreasing relatively slowly. When $w$ is larger than $1.9$, the community assignment which gives the highest $Q^{ud}$ is altered from a community assignment which is identical to the embedded community structure to a different community assignment. Because the nodes connected by the inter-ring links are more likely to be assigned into the same community in this new community assignment, the new assignment favors larger $w$ and $Q^{ud}$ starts to increase as $w$ becomes larger. Meanwhile, $Q^{lr}$ decreases continuously as $w$ increases, which is consistent with the fact that the community structure becomes weaker when $w$ gets larger.

Both the results of VOI and modularity show that our method can correctly and robustly detect the community structure of this model network, while the direction-ignoring method cannot. We also performed the same analysis to the model network of various values for $m$ and $n$. All the results are qualitatively the same with the result of $m=n=8$.

\section{Summary}

In this paper, we have presented a new definition of the modularity in directed networks by introducing a new quantity LinkRank, which indicates the importance of links in directed networks. The new modularity is related to the random walk process in the network, and the global meaning of our new modularity is the fraction of time spent moving within communities by a random walker minus the expected value of this fraction. And locally, the meaning of the new modularity is that a link with higher LinkRank is more likely to be assigned as an intra-community link than a link with lower LinkRank. The definition of community is also changed, according to the change of modularity. In this new definition, a community is a group of nodes in which a random walker is more likely to stay.

It has been proven that our new modularity is consistent
with the old modularity proposed by Newman and Girvan~\cite{Newman2004}. Also, other
methods of community identification are compared with our method. It is shown
that the method
proposed by Rosvall and Bergstrom~\cite{Rosvall2007} and our method share the same concept of
community structure in directed networks. A model network is designed to verify our method, and this model network can be used as a benchmark network in further studies of community identification. As most of the modularity optimization methods in undirected methods can be
applied to the directed networks by optimizing our new modularity, our method
would be very practical to use to identify communities in directed networks.

\begin{acknowledgments}
We thank J. D. Noh, P.-J. Kim, H. Youn, C. Jeon, N. Masuda, and S. Fortunato for their useful discussions. This work was supported by the Korean Systems Biology Program from the Ministry of Education, Science and Technology through the Korea Science and Engineering Foundation Grant No. M10309020000-03B5002-00000 (H.J. and Y.K.) and by Korea Research Council of Fundamental Science and Technology (S.S.).
\end{acknowledgments}

\end{document}